\documentclass[aps,prc,preprintnumbers,twocolumn,nofootinbib,superscriptaddress,showpacs]{revtex4-1} 
\usepackage{graphicx,amsmath,amssymb,bm,color}
\usepackage{enumerate}
\usepackage{multirow,dcolumn}

\allowdisplaybreaks[1]

\newcommand{\re}{r_\text{e}}
\renewcommand{\Re}{\text{Re}\,}

\newcommand{\ket}[1]{\left|#1\rangle\right.}
\newcommand{\bra}[1]{\langle#1|}
\newcommand{\braket}[2]{\langle#1\vphantom{#2}|#2\vphantom{#1}\rangle}

\newcommand{\matrixel}[3]
{\langle#1\vphantom{#2#3}|#2|#3\vphantom{#1#2}\rangle}

\renewcommand{\vec}[1]{\mathbf{#1}}

\newcommand{\up}{\uparrow}
\newcommand{\down}{\downarrow}

\begin{document}

%\preprint{INT-PUB-16-010}

\title{Quantum Monte Carlo calculations of two neutrons in finite volume}

\author{P.\ Klos}
\email[E-mail:~]{pklos@theorie.ikp.physik.tu-darmstadt.de}
\affiliation{Institut f\"ur Kernphysik, 
Technische Universit\"at Darmstadt, 
64289 Darmstadt, Germany}
\affiliation{ExtreMe Matter Institute EMMI, 
GSI Helmholtzzentrum f\"ur Schwerionenforschung GmbH, 
64291 Darmstadt, Germany}
\author{J.\ E.\ Lynn}
\affiliation{Institut f\"ur Kernphysik, 
Technische Universit\"at Darmstadt, 
64289 Darmstadt, Germany}
\affiliation{ExtreMe Matter Institute EMMI, 
GSI Helmholtzzentrum f\"ur Schwerionenforschung GmbH, 
64291 Darmstadt, Germany}
\author{I.\ Tews}
\affiliation{Institute for Nuclear Theory, University of Washington, Seattle, WA 98195-1550, USA }
\author{S.\ Gandolfi}
\affiliation{Theoretical Division, Los Alamos National Laboratory, Los Alamos, NM 87545, USA}
\author{A.\ Gezerlis}
\affiliation{Department of Physics, University of Guelph, Guelph, Ontario N1G 2W1, Canada}
\author{H.-W.\ Hammer}
\affiliation{Institut f\"ur Kernphysik, 
Technische Universit\"at Darmstadt, 
64289 Darmstadt, Germany}
\affiliation{ExtreMe Matter Institute EMMI, 
GSI Helmholtzzentrum f\"ur Schwerionenforschung GmbH, 
64291 Darmstadt, Germany}
\author{M.\ Hoferichter}
\affiliation{Institute for Nuclear Theory, University of Washington, Seattle, WA 98195-1550, USA }
\author{A.\ Schwenk}
\affiliation{Institut f\"ur Kernphysik, 
Technische Universit\"at Darmstadt, 
64289 Darmstadt, Germany}
\affiliation{ExtreMe Matter Institute EMMI, 
GSI Helmholtzzentrum f\"ur Schwerionenforschung GmbH, 
64291 Darmstadt, Germany}

\begin{abstract}

{\it Ab initio} calculations provide direct access to the properties of pure
neutron systems that are challenging to study experimentally.  In
addition to their importance for fundamental physics, their properties
are required as input for effective field theories of the strong
interaction. In this work, we perform auxiliary-field diffusion Monte
Carlo calculations of the ground and first excited state of two
neutrons in a finite box, considering a simple contact potential as
well as chiral effective field theory interactions. We compare the
results against exact diagonalizations and present a detailed analysis
of the finite-volume effects, whose understanding is crucial for
determining observables from the calculated energies. Using the
L\"uscher formula, we extract the low-energy $S$-wave scattering
parameters from ground- and excited-state energies for different box
sizes.

\end{abstract}

%\pacs{21.30.-x, 21.45.Bc, 21.60.Ka, 12.38.Gc}

\maketitle

\section{Introduction}

A major challenge of nuclear physics is the direct calculation of
nuclear observables from the underlying theory of quantum
chromodynamics (QCD). The most promising strategy today towards this
goal involves lattice QCD calculations for small nuclei and a sequence
of matching steps to advanced few- and many-body methods based on
effective field theories (EFTs) for nuclear forces.

Solving QCD on a lattice of discretized spacetime in Euclidean space
represents the only method to calculate nuclear observables from the
QCD Lagrangian~\cite{Bean10NPLQCD}.  Numerical calculations are only
feasible in a finite box where the energy spectrum is discrete.  Below
the first inelastic threshold, L\"uscher~\cite{Lues86long,Lues91Torus}
established a direct connection between the scattering phase shifts in
the infinite volume and finite-volume energy levels for two-particle
scattering. The extension of this connection to the three-particle
sector is the subject of ongoing research (see, for example,
Refs.~\cite{Kreuzer:2010ti,Polejaeva:2012ut,Briceno:2012rv,Hansen:2015zga}).
As a consequence, the scattering parameters, such as the scattering
length and effective range, can be extracted from finite-volume energy
levels obtained from lattice QCD.  However, as the L\"uscher result
corresponds to an EFT in which particles interact only via contact
interactions~\cite{Bean04TwoNLat}, it is not applicable to the regime
of nuclear physics where the nonanalyticities from pion exchange
become important, i.e., where exponential corrections start to set in.

Chiral EFT offers a framework to include explicit pions into the EFT
for low-energy nucleons~\cite{Wein90NFch,Wein91chNp}.  It has been
very successful in the past years and is used extensively to calculate
the properties of nuclei, electroweak transitions, and matter under
extreme conditions~\cite{Bedaque:2002mn,Epel09RMP,Machleidt:2011zz,%
Hammer:2012id,Holt2013PPNP,Hebe2015ARNPS}.  Chiral EFT is based on the
chiral symmetry of QCD and provides a systematic framework for nuclear
forces; in particular it predicts a hierarchy between two- and
multi-nucleon forces as well as external currents. It is therefore
possible to obtain a generalization of L\"uscher's result that
includes pions by matching finite-volume results from chiral EFT
interactions to lattice QCD calculations. This matching to chiral EFT
is done through adjustment of so-called low-energy constants (LECs),
which incorporate the degrees of freedom that were integrated out.
Once the LECs are determined, one is able to calculate phase shifts in
the infinite volume using few- and many-body methods to solve the
Schr\"odinger equation.  In this way, scattering parameters in the
infinite volume can be obtained also from finite-volume calculations
in smaller boxes where pion exchanges become relevant.

Presently, the LECs in chiral EFT are fit to experimental data, a
strategy that fails in channels where experimental data are scarce or
even nonexistent, such as the three-neutron system.  A future
alternative strategy consists of applying LECs matched to lattice QCD
to calculate other observables, which would provide a fully QCD-based
prediction for these observables.  At present, lattice QCD
calculations for systems with more than one nucleon are only available
for nonphysical quark masses. However, if the quark masses are within
the range of applicability of the chiral expansion, chiral EFT also
allows one to extrapolate to physical values and, in the process,
determine the LECs that govern the quark mass dependence, which are
largely unknown to date.

In this work, we perform finite-volume calculations of two particles
with chiral EFT interactions matched to experimental data. We focus on
the neutron-neutron ($nn$) system, motivated also by the fact that the
corresponding scattering length cannot be measured directly and
controversial data from indirect measurements would make an
independent {\it ab initio} verification particularly
valuable~\cite{Gardestig:2009ya}.
Our long-term goal 
is to establish a technique to match chiral nuclear interactions to
data from lattice QCD. This would allow the extraction of the LECs which
appear in chiral EFT directly from the lattice and thus facilitate a
path to calculate nuclear observables of larger nuclei directly from QCD.
To set the foundation for future work towards matching two- and 
higher-body systems, we require a numerical method that is easily
scalable to many particles and 
is well suited to studies in periodic boundary conditions.
We use the auxiliary-field diffusion
Monte Carlo (AFDMC) method, since it represents a very accurate method (see,
for example, Refs.~\cite{Carl15RMP,Gand2015ARNPS} for recent reviews),
for which calculations with chiral EFT interactions have become possible
only recently~\cite{Geze13QMCchi,Geze14long,Lynn14QMCln,Tews20163N,%
Lynn20163N}. Furthermore, performing finite-volume calculations is
straightforward in Quantum Monte Carlo (QMC) as coordinate space is
intrinsically constrained. A similar study on how to
exploit input from lattice QCD for determining LECs in pionless EFT
using AFDMC calculations can be found in Refs.~\cite{Barn2013uqa,Kirs2015lln}.

In this paper, we benchmark QMC calculations in finite volume
for ground and excited states, both for a contact potential as well as chiral
EFT interactions. In particular, we verify that QMC
finite-volume calculations, by means of the L\"uscher formalism,
reproduce the low-energy effective-range parameters corresponding to a
given $nn$ potential.
We thereby demonstrate that such QMC calculations
provide a reliable tool to establish a bridge between lattice QCD
calculations and chiral EFT, in particular in kinematic configurations
where the consideration of pion exchanges becomes mandatory and the
L\"uscher formula cannot be applied straightforwardly.

Generalizations of the L\"uscher formalism to multi-body systems prove
to be fairly
complex~\cite{Kreuzer:2010ti,Polejaeva:2012ut,Briceno:2012rv,Hansen:2015zga},
to the point that alternative strategies to extract infinite-volume
physics from finite-volume energy levels might be welcome. For QMC
methods to contribute in this direction, especially for channels where
resonances may occur, it is crucial that excited states can also be
accessed, in order to be able to identify the expected avoided level
crossing~\cite{Luscher:1991cf,Bernard:2007cm}. 
This presents a challenge.  Quantum Monte Carlo methods were
developed to solve the many-body Schr\"odinger equation of a given
system and find the lowest-energy state. As this particular state is given by the bosonic solution, nodal surfaces in the many-body wave function have to be introduced, something which can only be done approximately~\cite{Foul2001qmc,Carl15RMP,Piep04QMCex}.
While an exact solution to this problem is therefore not available at
the moment, we propose a strategy to obtain an approximate numerical
solution for the excited state. Although the $nn$ system strictly
speaking does not exhibit a resonance, but only a virtual state, the
$nn$ calculations presented in this paper can be considered a first
step towards this application.

This paper is organized as follows. A brief summary of the L\"uscher
formula is presented in Sec.~\ref{Sec_Luescher}. Section~\ref{Sec_QMC}
gives an overview of the AFDMC method for ground and excited states. A
special emphasis is placed on the construction of trial wave functions
that become important for the calculation of excited states. In
Sec.~\ref{Sec_results}, we present AFDMC results for finite-volume
calculations of $nn$ energies for a contact potential as well as
chiral EFT interactions at different orders. Both ground states and
excited states are compared with results from the L\"uscher
formula and also to exact
diagonalizations. Based on this, we analyze in detail the
finite-volume effects and deviations caused by pion exchanges. We then
successfully extract scattering parameters from the finite-volume
results using the L\"uscher formula.  Finally, we conclude in
Sec.~\ref{Sec_conclusions}.

\section{L\"uscher formula}
\label{Sec_Luescher}

The very low-momentum properties of nucleon-nucleon (NN)
interactions can be efficiently described within a pionless
EFT~\cite{Kaplan:1998we,vanKolck:1998bw,Chen:1999tn}. Constrained to
short-range interactions the Lagrangian becomes a series of local
operators that consist of derivatives acting on nucleon
fields. Applying dimensional regularization with power-divergence
subtraction~\cite{Kaplan:1998we}, the scattering amplitude for
two-body elastic scattering can be written in terms of a single scalar
integral, whose divergent part becomes absorbed into the
renormalization of the LECs of the theory. In a box of size $L^3$ with
periodic boundary conditions it is then possible to relate the energy
eigenstates of the two-body system to the $S$-wave phase shift
$\delta_0(p)$ in infinite volume. The eigenvalues for the energy
$E=p^2/M$, with relative momentum $p$ and nucleon mass $M$, are given
in terms of solutions of the L\"uscher
formula~\cite{Lues86long,Lues91Torus}\footnote{This version only holds for an $S$-wave projected potential, with corrections entering
at the level of $G$-waves. We considered the corresponding generalized relation~\cite{Lues86long,Lues91Torus} as well, but found the corrections
to be negligible due to the large suppression of the physical $G$-wave phase shift.}
\begin{equation}
p\cot \delta_0(p)=\frac{1}{\pi L}S\biggl(\Bigl(\frac{Lp}{2\pi}\Bigr)^2\biggr) \,.
\label{Eq_Luescher}
\end{equation}
$S(\eta)$ can be defined as a regulated sum
\begin{equation}
\label{S_eta_loop}
S(\eta)=\lim\limits_{\Lambda \rightarrow \infty}\Biggl(\sum_{|{\bf j}|<\Lambda}\frac{1}{{\bf j}^2-\eta}-4\pi\Lambda\Biggr) \,,
\end{equation}
which runs over all three-vectors of integers ${\bf j}$ with
$|{\bf j}|<\Lambda$.  A more detailed discussion of $S(\eta)$ as well
as its practical implementation for a numerical evaluation are
summarized in Appendix~\ref{app:Seta}.

This form of the L\"uscher formula emerges naturally in pionless EFT
when the loop integral is replaced by a discrete sum over the momentum
states allowed on the lattice~\cite{Bean04TwoNLat}. In this
derivation, Eq.~\eqref{Eq_Luescher} strictly holds as long as a
description in pionless EFT is justified. Due to the $t$-channel cut
in the one-pion exchange, this restricts its range of validity to
$|p|<m_\pi/2$ in the complex $p$ plane.  However, as shown
in Refs.~\cite{Lues86long,Lues91Torus}, corrections to
Eq.~\eqref{Eq_Luescher} for momenta below the first inelastic
threshold $|p|<\sqrt{m_\pi M}$ are suppressed by $e^{-m_\pi L}$, so that
in practice the relation can be used as long as $m_\pi L$ is
sufficiently large. In Ref.~\cite{Sato:2007ms}, the size of these corrections in the two-nucleon system
was estimated for EFT-inspired potentials with
pion exchange and contact interactions.

For low-energy $NN$ scattering the first parameters of the
effective-range expansion, the scattering length $a$ and the effective
range $\re$, are sufficient for an accurate description of the phase
shift:
\begin{equation}
p\cot \delta_0(p)=-\frac{1}{a}+\frac{1}{2}\re p^2+\mathcal{O}(p^4) \,.
\label{Eq:effRexp}
\end{equation}
Therefore, it is possible for a given scattering length and effective
range to predict the energies of ground and excited states in a finite
volume.  Vice versa, given a set of data points $\{E_i,\Delta E_i\}$
for the energy eigenvalues for different box sizes $L_i$ one can
determine the scattering parameters $a$ and $\re$ that best fulfill
Eq.~\eqref{Eq_Luescher}, where the left-hand side has been replaced by
Eq.~\eqref{Eq:effRexp}. In fact, whenever energy levels become
negative, Eq.~\eqref{Eq_Luescher} provides a constraint in the
unphysical region that cannot immediately be translated into a
corresponding value for the phase shift. In such cases, the
effective-range expansion~\eqref{Eq:effRexp}, in addition to providing
a convenient parametrization of the phase shift, serves another
purpose, namely that of stabilizing the analytic continuation towards
the physical region, which can only be performed if the functional
form is known.  This situation is realized for the ground-state energy
of the two-neutron system.

Although Eq.~\eqref{Eq_Luescher} could still be used for
$m_\pi/2<|p|<\sqrt{m_\pi M}$ provided the volume is sufficiently
large, the validity of the analytic continuation based on the
effective-range expansion~\eqref{Eq:effRexp} is limited by the
$t$-channel pion exchange, which in the partial-wave projection
generates cuts on the imaginary momentum axis starting at $p=\pm i
m_\pi/2$.  Therefore, if points with $|p|>m_\pi/2$ were to be
included, these cuts would have to be accounted for explicitly in the
functional form used in the analytic continuation.  For this reason we
restrict all fits in this paper to points within the strict radius of
convergence of pionless EFT $|p|<m_\pi/2$.
 
Since $S(\eta)$ is not invertible, it is not possible to directly
define a function $E=E(L,a,\re)$ which could be used in a standard
$\chi^2$ fit, so that we minimize instead
\begin{equation}
\chi^2=\sum_{i=1}^N\frac{\biggl(\frac{1}{a}-\frac{1}{2}\re M E_i+\frac{1}{\pi L_i}S\Bigl(\bigl(\frac{L_i}{2\pi}\bigr)^2M E_i\Bigr)\biggr)^2}{\sigma^2_i} \,,
\label{Eq:chi2}
\end{equation}
with standard deviations obtained from Gaussian error propagation
\begin{equation}
\sigma^2_i=\biggl[-\frac{1}{2}\re M+\frac{M L_i}{4\pi^3}S'\biggl(\Bigl(\frac{L_i}{2\pi}\Bigr)^2M E_i\biggr)\biggr]^2(\Delta E_i)^2 \,,
\end{equation}
and $S'(\eta)=dS(\eta)/d\eta$. Parameter errors are estimated from the Hessian
\begin{equation}
H=\frac{1}{2}\left(
\begin{array}{cc}
\frac{\partial^2 \chi^2}{\partial a^2}&\frac{\partial^2 \chi^2}{\partial a \partial \re}\\
\frac{\partial^2 \chi^2}{\partial \re \partial a}&\frac{\partial^2 \chi^2}{\partial \re^2}
\end{array}
\right)\biggr|_{a_\text{min}, (\re)_\text{min} } \,,
\end{equation}
according to
\begin{equation}
\Delta a=\sqrt{(H^{-1})_{11}} \,, \qquad \Delta \re=\sqrt{(H^{-1})_{22}} \,.
\end{equation}
Based on these equations we demonstrate the feasibility of an
extraction of scattering parameters from finite-volume QMC
calculations in Sec.~\ref{Sec_results}, for both a contact potential
as well as chiral EFT interactions.

\section{Quantum Monte Carlo}
\label{Sec_QMC}

The AFDMC method has been successfully applied to both homogeneous and
inhomogeneous neutron matter in the past decade (see
Refs.~\cite{Carl15RMP,Gand2015ARNPS} for a summary of results and a
more detailed description of the method) and more recently has shown
promising progress towards generalization to nuclear matter and
nuclei~\cite{Gandolfi:2014ewa}.  In this section we review the basic
concepts of the AFDMC method and how it was applied in the two-neutron
system.  We give particular attention to the calculation of excited
states, which is in general a nontrivial task for QMC methods.

The aim of diffusion Monte Carlo methods is to solve the many-body
Schr\"odinger equation by means of stochastically projecting out
the lowest-energy state from a given trial wave function $\psi_T$,
\begin{align}
\psi(\tau)=e^{-(H-E_T)\tau}\psi_T \,,
\label{eq:dmc}
\end{align}
where the trial energy $E_T$ is a constant that is used to control the
normalization.  In the limit of large imaginary time $\tau=it$, only
the lowest-energy state not orthogonal to $\psi_T$ survives:
$\psi(\tau\rightarrow\infty)\rightarrow\psi_0$.  Therefore, the choice
of a trial wave function with symmetries appropriate to the state
under study is an important consideration, a point to which we return
when discussing the calculation of excited states.  For strongly
interacting many-body systems it is not possible to calculate directly
the imaginary-time propagator
\begin{equation}
G_{\alpha\beta}(\vec{R},\vec{R}^\prime;\tau)=
\matrixel{\alpha,\vec{R}}{e^{-(H-E_T)\tau}}{\beta,\vec{R}^\prime} \,,
\end{equation}
where $\vec{R}=\left\{\vec{r}_1,\vec{r}_2,\ldots,\vec{r}_A\right\}$ is
the configuration vector of the $A$ nucleons and $\alpha,\beta$ are
spin-isospin indices.  However, in the small imaginary-time limit, the
calculation is tractable and the propagation is performed as a
sequence of small-time evolutions.  Then, the realization of
Eq.~\eqref{eq:dmc} is given by the path integral (omitting the
spin-isospin labels)
\begin{equation}
\psi(\vec{R}_N,\tau)=\int\prod_{i=0}^{N-1}d\vec{R}_i\,
G(\vec{R}_{i+1},\vec{R}_i;\Delta\tau)\psi_T(\vec{R}_0) \,,
\end{equation}
where $\Delta\tau$ is the small imaginary time step and the paths
$\vec{R}_i$ are sampled by Monte Carlo.

The AFDMC method takes as a basis state the tensor product of the $3A$
coordinates of the $A$ nucleons and the tensor product of the four
complex amplitudes for each nucleon to be in a state
$\ket{s}=\ket{p\up,p\down,n\up, n\down}$.  That is,
\begin{equation}
\ket{\vec{R}S}=\ket{\vec{r}_1s_1}\otimes\ket{\vec{r}_2s_2}\otimes\cdots
\otimes\ket{\vec{r}_As_A} \,.
\end{equation}
As a consequence of the choice of basis, the propagator must contain,
at most, linear operators in spin-isospin space.  Therefore, the
Hubbard-Stratonovich transformation is used to linearize quadratic
operators in the Hamiltonian
\begin{equation}
e^{O^2/2}=\frac{1}{\sqrt{2\pi}}\int_{-\infty}^\infty dx \,
e^{-x^2/2}e^{xO}\,,
\end{equation}
introducing the auxiliary fields $x$, which are Monte Carlo sampled to
perform the integrals.
This choice provides for polynomial scaling with nucleon number, because the
spin-isospin states are sampled instead of summed explicitly as in, for
example, the Green's function Monte Carlo (GFMC) method~\cite{Carl15RMP}.

In this work, we take a Jastrow trial wave function, which is a
product of central correlations acting on a Slater determinant of
single-particle orbitals,
\begin{equation}
\ket{\psi_J}=\Big[\prod_{i<j}f^{c}(r_{ij})\Big]\ket{\Phi} \,,
\end{equation}
with
$\braket{\vec{R}S}{\Phi}=\mathcal{A}[\braket{\vec{r}_1s_1}{\phi_1}\cdots
\braket{\vec{r}_2s_2}{\phi_2}\cdots\braket{\vec{r}_As_A}{\phi_A}]$.
The Jastrow wave function incorporates the dominant short-range
central correlations into the wave function, by a solution of the
radial Schr\"odinger equation in the given spin-isospin channel of the
Hamiltonian.  For neutrons in a cubic periodic box of volume $L^3$,
the single-particle orbitals are taken as plane waves:
$\phi_\alpha(\vec{r}_i,s_i)=
e^{i\vec{k}_\alpha\cdot\vec{r}_i}\chi_{s,m_s}(s_i)$, with
$\vec{k}_\alpha=\frac{2\pi}{L}\vec{n}_\alpha$ and $\vec{n}_\alpha$ being a
vector of integers. $\chi_{s,m_s}$ denotes the spin eigenstates.  For
two neutrons, only the lowest two states with
$\vec{k}_1=\vec{k}_2=\vec{0}$ are occupied, leaving the Slater
determinant $\braket{\vec{R}S}{\Phi}$ independent of spatial
coordinates.

Imposing periodic boundary conditions is equivalent to identifying the
endpoints of each Cartesian interval.
This implies that, in coordinate space, the potential includes,
in addition to the original potential $V(\vec{r})$, copies from the surrounding boxes
\begin{equation}
\label{eq:imgsum}
V(\vec{r})\to\sum_{\vec{n}\in\mathbb{Z}^3}V(\vec{r}+\vec{n}L)
\end{equation}
to preserve periodicity~\cite{Lues86long}.
As long as the range of the potential, characterized by the effective
range $\re$ for example, is small compared to the box size,
$\re\ll L$, the higher terms in the sum in Eq.~\eqref{eq:imgsum} can be safely
ignored.
However, when the box size becomes comparable to the range of the
potential $L\sim \re$, these higher terms need to be included
in both the expectation value of the
Hamiltonian and in the calculation of the propagator in order to
maintain the periodic boundary conditions.
Below, if necessary, we consider terms corresponding up to either the nearest, second-to-nearest, or third-to-nearest boxes around the original one and thereby check for convergence of the sum in Eq.~\eqref{eq:imgsum}.

The calculation of excited states can be a challenging task for
diffusion Monte Carlo methods.  Since such methods always project, out
of a trial wave function, the lowest-energy state of a given
Hamiltonian, care must be taken to ensure orthogonality to, for
example, the ground state (if the first-excited-state solution is
sought).  For nuclei, in many cases, the excited state which is
desired has quantum numbers distinct from the ground state.  In this
case, all that is required is to construct a trial wave function with
the appropriate quantum numbers~\cite{Pudliner:1995wk}.  However, in
some cases, for example the Hoyle state of $^{12}$C, the desired
excited state has the same quantum numbers as the ground state, and
then more care in constructing an appropriate trial wave function is
required~\cite{Piep04QMCex}.  In the two-neutron system for
low-energy scattering, we consider only the case where the neutrons
are in a singlet spin state ($^1S_0$), which corresponds to the state
described by the L\"uscher formula in Sec.~\ref{Sec_Luescher}, and
therefore the excited states possess the same quantum numbers as the
ground state.  Such excited scattering states have not been calculated
previously using the AFDMC method.

The trial wave function for the first excited state calculated here
was determined as follows.  We assume a nodal surface defined by a
particular relative distance $r_\text{node}$ between the two
particles.  Since our Slater determinant is spatially independent, we
introduce the node in the central correlation of the Jastrow wave
function such that $\psi_J(r_\text{node})=0$.  This implies that there
is no angular dependence and the nodal surface is a sphere in relative
coordinates.  The validity of this assumption and an estimate for the
related systematic error as well as an improved nodal surface will be discussed later. To determine the
nodal position, we adopt the iterative approach described below.

Quantum Monte Carlo methods typically require local potentials.  For
an eigenstate of a local Hamiltonian, the solution of the
Schr\"odinger equation must yield the same energy independent of the
coordinates at which it is evaluated.  Evaluating $[H\psi({\bf
R})]/\psi({\bf R})$, where $\psi$ is the exact solution of the
problem, should therefore yield the same energy for a configuration of
the two particles ${\bf R}=\{{\bf r}_1,{\bf r}_2\}$ with relative
distance $r<r_\text{node}$ or $r>r_\text{node}$.  As a consequence,
the node position can be obtained by performing separate AFDMC
simulations in the two subspaces divided by the nodal surface and
adjusting the node position such that the AFDMC energies in the two
subspaces agree.  Each of these simulations starts from initial
configurations where all walkers are placed in one of the two
subspaces.  For an arbitrarily chosen node position the two
simulations will yield different energies.  Moving the node position
in the relative coordinate such that the two independent simulations
yield the same energy within statistical uncertainties leads to the
results presented in the next section.  As the constrained-path
approximation~\cite{Zhang:2003zzk}, which we use to tame the sign
problem, prohibits walkers from crossing the nodal surface, this is
equivalent to performing a simulation in a space which is limited to
the region where the trial wave function does not change sign.

\section{Results}
\label{Sec_results}

We perform AFDMC simulations of two neutrons in a cubic box with
periodic boundary conditions for both a simple contact potential as
well as chiral EFT interactions.  Both ground-state and
first-excited-state energies are calculated and compared to exact
solutions derived from the L\"uscher formula in
Eq.~\eqref{Eq_Luescher} with the effective range expansion in
Eq.~\eqref{Eq:effRexp}.  The box size was varied from $L=5$~fm to
$L=50$~fm.

\subsection{Contact interaction}
\label{Sec:trialpot}

\begin{figure}[t]
\includegraphics[width=0.48\textwidth,clip=]{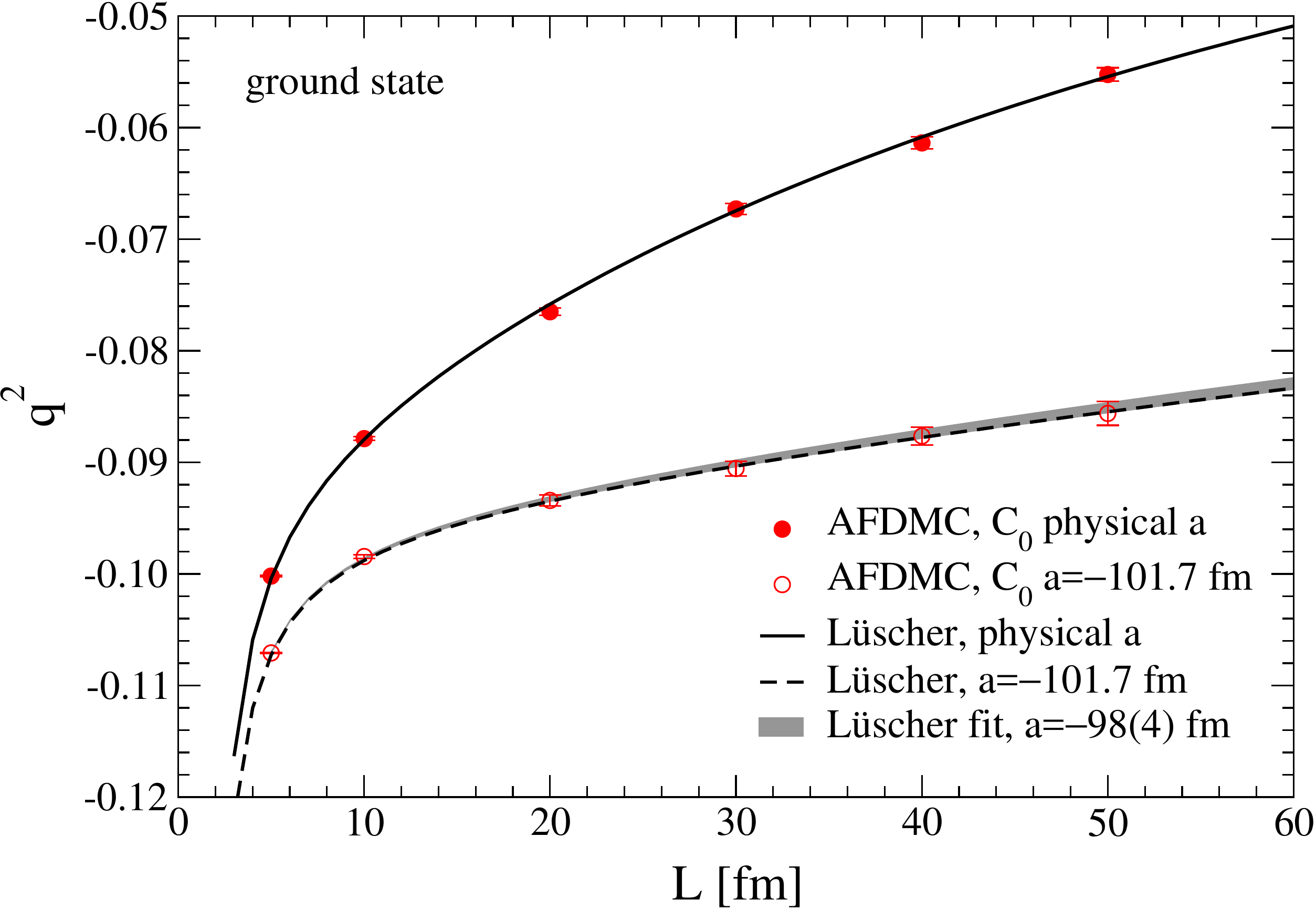}
\caption{(Color online) AFDMC results for the energy of two neutrons 
in the ground state in finite volume with the contact
potential~\eqref{Eq_testpot} for different box sizes~$L$ compared with
the L\"uscher formula.  $C_0$ is adjusted to give the physical $nn$
scattering length $a=-18.9$~fm (closed circles and solid line) and to give
a very large scattering length $a=-101.7$~fm (open circles and dashed
line).  The gray band shows a fit (as described in the text) to the
AFDMC results for $a=-101.7$~fm.  The energies are given in terms of
the dimensionless quantity $q^2=E ML^2/(4\pi^2)$.}
\label{Fig_TP_GS}
\end{figure}

First, we consider a contact interaction independent of spin and
isospin operators smeared out by a regulating function
$V(r)=C_0\delta(r)\rightarrow C_0\delta_{R_0}(r)$ with
\begin{align}
\delta_{R_0}(r)= \frac{1}{\pi\Gamma(3/4)R_0^3}
\exp\biggl[-\Bigl(\frac{r}{R_0}\Bigr)^4\biggr] \,,
\label{Eq_testpot}
\end{align}
where $C_0$ is a constant and $R_0=1.0$~fm determines the range of the
regulator~\cite{Geze14long}.

This potential corresponds to the smeared-out contact interaction
which is used in the local chiral EFT interactions considered in
Sec.~\ref{Sec:chiralInt}.  Furthermore, up to the regulator, this
potential corresponds to the interaction underlying the derivation of
the L\"uscher formula described in Sec.~\ref{Sec_Luescher}.  For our
calculations we take $L>R_0$ to minimize the finite-cutoff effects.
Results obtained using this potential will serve as a benchmark for
the AFDMC method in the two-particle system since we expect agreement
with the L\"uscher prediction up to statistical uncertainties.

\begin{figure}[t]
\includegraphics[width=0.48\textwidth,clip=]{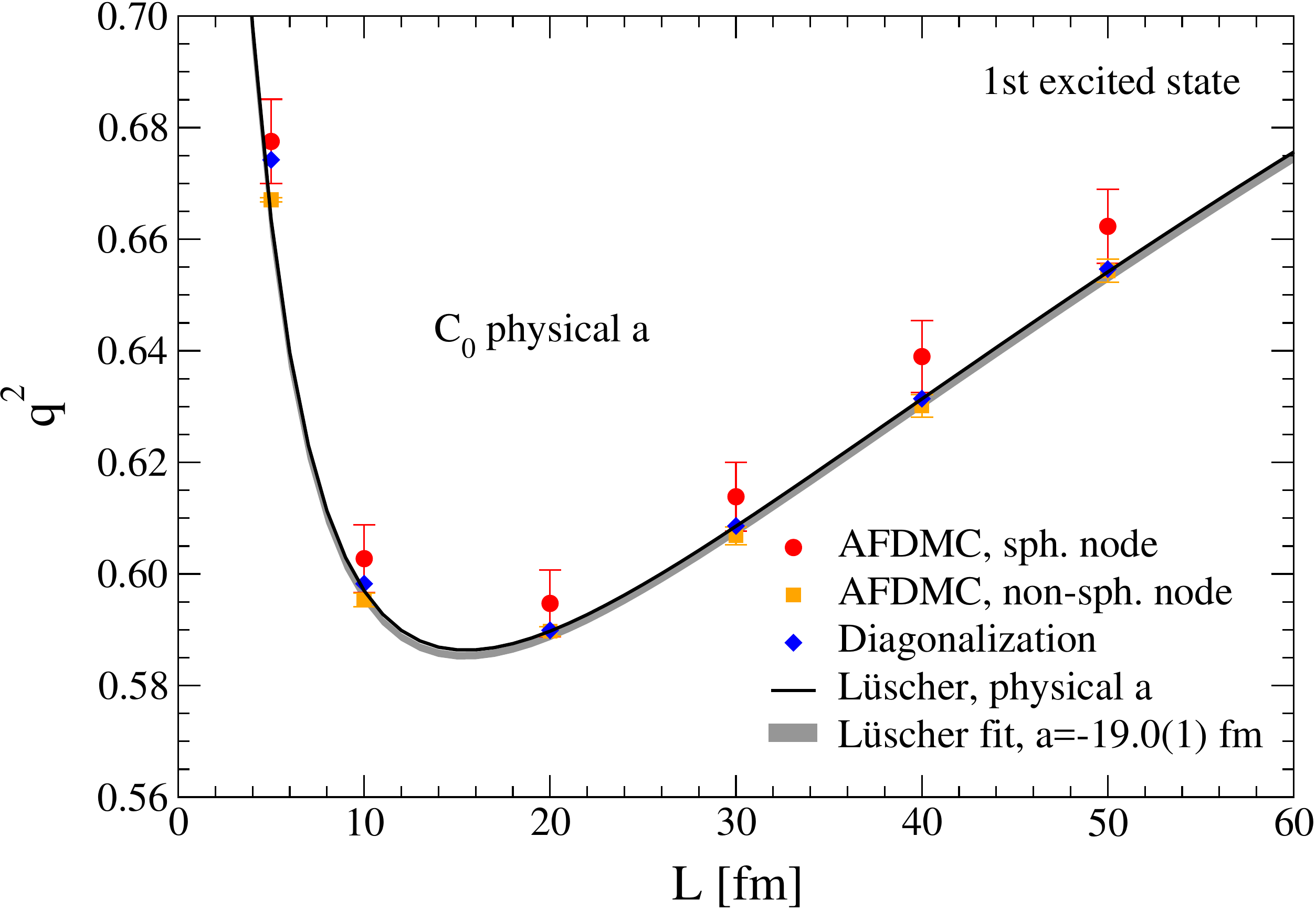}
\caption{(Color online) AFDMC results for the energy of two neutrons in 
the first excited state in finite volume with the
potential~\eqref{Eq_testpot} for different box sizes~$L$ (red circles)
compared with the L\"uscher formula (solid line).  The error bars on
the AFDMC results with a spherical nodal surface include both statistical uncertainties and a
systematic uncertainty of 1\% discussed in the text in
Sec.~\ref{Sec:Diag}.  $C_0$ is adjusted to give the physical $nn$
scattering length $a=-18.9$~fm.  The energies are given in terms of
the dimensionless quantity $q^2=E ML^2/(4\pi^2)$.  Also shown are the
energies calculated by exact diagonalization (blue diamonds) as
discussed in Sec.~\ref{Sec:Diag}.}
\label{Fig_TP_EX}
\end{figure}

Table~\ref{tab:gfmcafdmc} compares results of both the AFDMC and GFMC methods for representative box sizes. The GFMC calculations reproduce the AFDMC results within uncertainties, i.e., using the same number of configurations the uncertainties in the two methods are similar. Therefore, results will be shown only for the AFDMC method. Concerning the convergence of the sum in Eq.~\eqref{eq:imgsum}, we found for this potential that taking only the interaction in the original box into account is sufficient, which is consistent with the range of the potential.

\begin{table}[t]
\caption{Comparison of ground-state results for two different
potentials with both the AFDMC and GFMC methods for several box
sizes $L$.}
\label{tab:gfmcafdmc}
\begin{ruledtabular}
{\renewcommand{\arraystretch}{1.10}
\begin{tabular}{ccdd}
&&\multicolumn{2}{c}{$q^2$}\\
\cline{3-4}
Potential&$L\ (\text{fm})$&\multicolumn{1}{c}{$\text{AFDMC}$}&
\multicolumn{1}{c}{$\text{GFMC}$}\\
\hline
\multirow{3}{*}{$C_0$ physical $a$}
&5& -0.1001(3)&-0.0999(1)\\
&10&-0.0879(7) &-0.0875(4)\\
&20&-0.069(2)  &-0.072(2)\\
\\
\multirow{3}{*}{LO $R_0=1.0$ fm}
&5&-0.1179(6)&-0.1178(1)\\
&10&-0.0931(6)&-0.0940(4)\\
&20&-0.079(2)&-0.077(1)\\
\end{tabular}}
\end{ruledtabular}
\end{table}

Figure~\ref{Fig_TP_GS} compares the ground-state energies in terms of
the dimensionless quantity $q^2=E ML^2/(4\pi^2)$ obtained from AFDMC
simulations with the exact solutions from the L\"uscher formula for
two different sets of scattering parameters.  In the first case we
used $C_0=-2.2369$~fm$^2$, which corresponds to the physical value for
the $nn$ scattering length of $a=-18.9$~fm and an effective range
$\re=1.096$~fm in infinite volume.  The second case shows results for
a potential with $C_0=-2.319$~fm$^2$ corresponding to a very large
scattering length of $a=-101.7$~fm and an effective range of
$\re=1.074$~fm.  As can be seen, the agreement between the L\"uscher
results and the AFDMC simulations of the ground state is excellent
over the full range of box sizes considered.  It is worth pointing out
the precision possible with the AFDMC method even at the extremely low
densities of $2/(50\text{ fm})^3\sim n_0/10^4$, with $n_0$ being the
saturation density of nuclear matter.

In future applications, one could take finite-volume results from
lattice QCD calculations, extract scattering parameters from them, and
adjust LECs in chiral EFT interactions to match these scattering
parameters.  Here we demonstrate this idea by extracting the
scattering parameters from the AFDMC results in several cases.  We
propagate the estimated uncertainties from the AFDMC simulations
through the $\chi^2$ fit discussed in Sec.~\ref{Sec_Luescher} and fit
the first two or three parameters of the effective range expansion.
In particular, in order to see how robust the extraction of the
infinite-volume scattering parameters from the AFDMC calculations is,
we consider the contact potential with the very large scattering
length.  Here we performed a two-parameter fit to $a$ and $\re$ by using the ground-state data yielding $a=-98(4)$~fm and $\re=1.066(7)$~fm
which agree within the uncertainties with the infinite-volume
parameters given above.  Figure~\ref{Fig_TP_GS} shows the
corresponding L\"uscher result.  The large uncertainty in the fitted
scattering length of more than 4\% could be reduced significantly when
including more data at $L\geqslant 20$~fm where $a$ dominates the fit.

Results for excited-state energies of two neutrons with the contact
potential with physical scattering length are shown in
Fig.~\ref{Fig_TP_EX}.  AFDMC results are shown for both a spherical nodal surface as described in Sec.~\ref{Sec_QMC} and a nonspherical nodal surface, as will be introduced in Sec.~\ref{Sec:Diag}. The AFDMC results from the spherical node are systematically above the
L\"uscher results by $\sim 1\%$; however, the overall trend is
correctly reproduced.  The global deviations can be understood when
taking the assumption of a spherical nodal surface into account.  An
analysis of the systematic error related to this assumption will be
discussed in the following section. The results from the nonspherical nodal surface reproduce the L\"uscher results very accurately.

Figure~\ref{Fig_TP_EX} also shows
a fit to the combined data of all AFDMC results for ground and excited
states with the improved nodal surface for the contact potential with physical scattering length.
For the fit, the first three coefficients of the effective range expansion
including the shape parameter were taken into account. The sensitivity to the shape parameter is largest for the states with largest momentum $p$, which correspond to excited states for small box sizes. As there are only a few of these contained in the data set the
shape parameter cannot be determined with enough precision. The
three-parameter fit yields a reduced $\chi^2$ value of $0.74$, a
scattering length of $a=-19.0(1)$~fm, and an effective range of
$\re=1.081(5)$~fm. These both agree well with the infinite-volume
values given above.  A more detailed discussion covering different
aspects of extracting the scattering parameters from finite-volume
energies of ground and excited states is given in
Sec.~\ref{Sec:chiralInt}.

\subsection{Exact diagonalization and nodal surface}
\label{Sec:Diag}

Diffusion Monte Carlo simulations do not provide direct access to the
propagated wave function.  To study the nodal structure of
the wave function of two neutrons in a box we diagonalize the
Hamiltonian in an appropriate basis.  The computational effort can be
minimized by choosing basis states satisfying the boundary conditions
of the system under study.  We are interested in the
zero-total-momentum eigenstates of a cubic box with periodic boundary
conditions.  Furthermore, since we are limiting ourselves to $S$-wave
states we only need to take basis functions of even parity into
account.  A convenient set of basis functions meeting these
requirements is given by
\begin{align}
\psi^{3\text{D}}_{nmk}({\bf r})&=\psi_{n}(x)\psi_{m}(y)\psi_{k}(z) \,, \notag\\
\psi_{n}(x)&=\sqrt{\frac{2-\delta_{0n}}{L}}\cos\biggl(\frac{2\pi}{L}nx\biggr) \,,
\end{align}
where $n=0,1,2,\ldots$ and ${\bf r}={\bf r}_1-{\bf r}_2$.

\begin{figure}[t]
\includegraphics[width=0.48\textwidth,clip=]{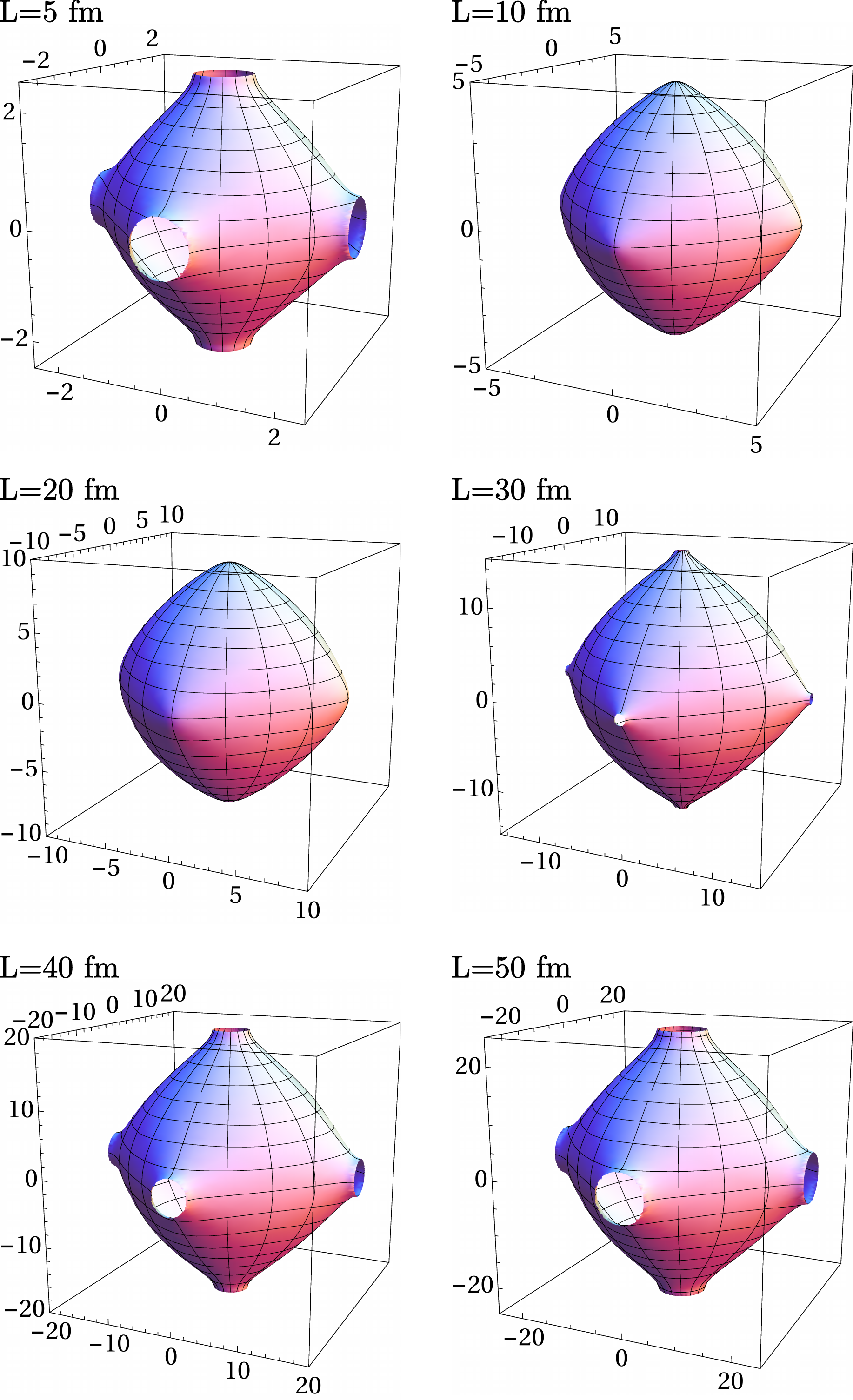}
\caption{(Color online) The nodal surfaces $r_\text{node}(\theta,\phi)$
of the first excited states of the contact
potential~\eqref{Eq_testpot} with the physical $nn$ scattering length
$a=-18.9$~fm for different box sizes~$L$.  The wave functions from
which the nodal surfaces are extracted are obtained via
diagonalization.  See text for details.}
\label{Fig_node}
\end{figure}

Exploiting further the cubic symmetry of the box, which implies that
the eigenstates have to remain invariant under exchange of coordinates
$x,y,z$, the number of basis states can be reduced by defining
symmetrized states for $n\leqslant m \leqslant k$:
\begin{align}
&\psi^{3\text{D sym}}_{nmk}(x,y,z)=
N\sum_{\{n,m,k\}}\psi^{3\text{D}}_{nmk}(x,y,z) \,,\notag\\
&N=\left\lbrace\begin{array}{ll}
1/\sqrt{6}&\text{for }n\neq m\neq k\\
1/\sqrt{12}&\text{for }n=m\neq k \\
1/6&\text{for }n=m=k\,,
\end{array}\right.
\label{Eq:diagbasis}
\end{align}
where the sum runs over all permutations of $\{n,m,k\}$.  The number
of basis states implicitly set by $n_\text{max}>n,m,k$ has to be
chosen such that the energy eigenstates are converged. As the box size
$L$ grows, $n_\text{max}$ has to be increased because higher momentum
states contribute to the eigenstates. The calculations for
$L=5,10,20,30,40,50$~fm presented here were performed by using
$n_\text{max}=10,16,32,48,54,54$, respectively.  Solving the
eigensystem $H\psi=E\psi$ yields the eigenstates $\psi_\text{gs}$ and
$\psi_\text{ex}$ corresponding to the ground- and first-excited-state
energies $E_0$ and $E_1$ in terms of the basis defined in
Eq.~\eqref{Eq:diagbasis},
\begin{equation}
\psi_\text{gs/ex}=\sum_{\substack{n,m,k<n_\text{max}\\n\leqslant m\leqslant k}}
c_{nmk}^\text{gs/ex} \, \psi^{3\text{D sym}}_{nmk} \,.
\end{equation}
The excited-state energies for the contact
potential~\eqref{Eq_testpot} with the physical $nn$ scattering length
$a=-18.9$~fm are shown in Fig.~\ref{Fig_TP_EX}. The results for the
excited state from the diagonalization agree within $0.01 \%$ for
$L=20,30,40$~fm with the exact results obtained from the L\"uscher
formula.  At $L=5$~fm ($L=10$~fm) a deviation of $1.6\%$ ($0.2\%$)
from the L\"uscher result can be observed.  For the small boxes,
especially for $L=5$~fm, the range of the potential $R_0$ and the box
size $L$ are of the same order and the finite range of the contact
potential~\eqref{Eq_testpot} becomes relevant.  Hence, a deviation
from the L\"uscher prediction is expected. For $L=50$~fm a deviation
of $\sim 0.1\%$ was obtained implying that $n_\text{max}$ needs to be
increased in order to reach convergence.

\begin{figure}[t]
\includegraphics[width=0.48\textwidth,clip=]{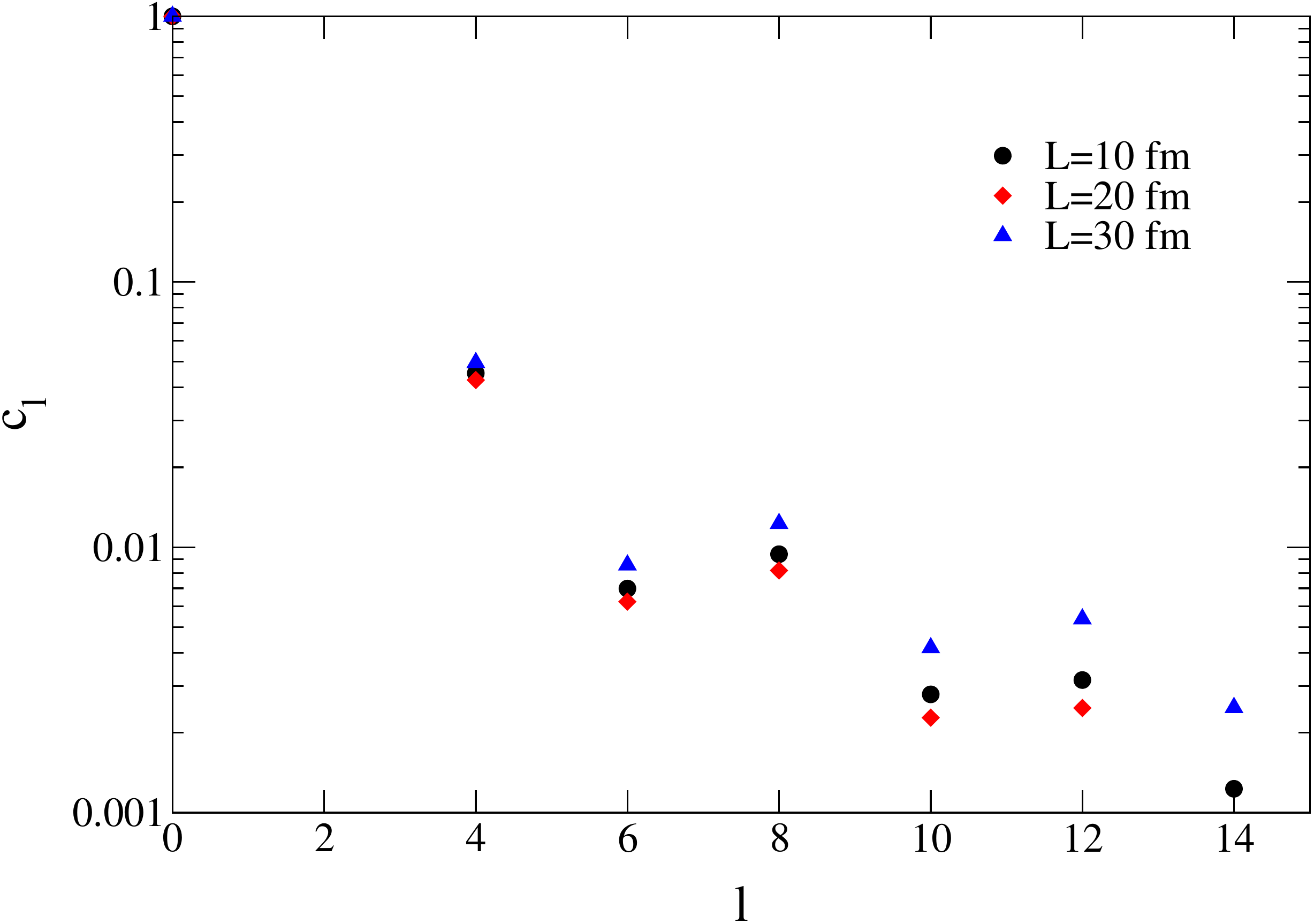}
\caption{(Color online) The coefficients $c_l$ of the decomposition 
of the nodal surfaces in terms of cubic harmonics $Y_l^c$ as in
Eq.~\eqref{Eq:Yldecomp} as a function of $l$.  The coefficients for
each box size shown are normalized such that $c_0=1$.  The $y$-axis is
log scaled showing the power-law suppression of the two coefficients
$c_4$ and $c_6$ with comparison to $c_0$.}
\label{Fig_Yldecomp}
\end{figure}

The nodal surface $r_\text{node}(\theta,\varphi)$ can be extracted
from the wave function of the excited state by solving
$\psi_\text{ex}(r_\text{node},\theta,\varphi)=0$ for
$\theta\in[0,\pi]$, $\varphi\in[0,2\pi]$.  Figure~\ref{Fig_node} shows
the nodal surfaces for different box sizes.  For the contact
potential~\eqref{Eq_testpot} with the physical $nn$ scattering length
$a=-18.9$~fm, the nodal surface is not spherical for any box size and
not even closed for box sizes of 5, 30, 40, and 50~fm.  To
estimate the systematic error caused by assuming spherical symmetry in
the nodal surface for the AFDMC simulations, we decomposed the nodal
surfaces in real spherical harmonics $Y_{lm}$:
\begin{align}
r_\text{node}(\theta,\phi)=\sum_{l=0}^\infty \sum_{m=-l}^{l} c_{lm} Y_{lm}(\theta,\phi) \,.
\end{align} 
We found that the coefficients $c_{lm}$ vanish for all $l, m$ other
than $l=0,4,6,8,...$ and $m=0,4,8,...$, which suggests that there is a
more appropriate set of functions in which one can expand the nodal
surface.  Indeed, in a cubic box the rotation symmetry group is broken
down to the cubic symmetry group $O_h$.  The irreducible
representation of $O_h$ is given by combinations of spherical
harmonics, so-called cubic harmonics $Y^c_l$, with
\begin{align}
Y^c_l(\theta,\phi)=\sum_{m=0,4,8,...}a_{lm}Y_{lm}(\theta,\phi) \,,
\end{align}
where the coefficients $a_{lm}$ are given in Ref.~\cite{Mugg72CubY}.
We found that for a given $l$ the coefficients
$c_l=\frac{c_{lm}}{a_{lm}}$ agree for all $|m|\leqslant l$. Hence, as
expected it is possible to expand the nodal surface in terms of cubic
harmonics,
\begin{align}
r_\text{node}(\theta,\phi)=\sum_l c_l Y^c_l(\theta,\phi) \,.
\label{Eq:Yldecomp}
\end{align}
The corresponding coefficients $c_l$ for box sizes where the nodal
surface is closed ($L=10,20$~fm) and almost closed ($L=30$~fm) are
shown in Fig.~\ref{Fig_Yldecomp}. Note that the coefficients are
normalized such that $c_0=1$. $c_0$ being much larger than $c_l$ for
$l\geqslant 4$ justifies the approximation of the nodal surface as a
sphere used in the AFDMC simulations because the spherical contribution
from $Y^c_0$ dominates the nodal surface.  We do not perform the
decomposition for the other box sizes since the holes in the surfaces
cause large uncertainties when decomposing into cubic harmonics.

The radial solution of the two-particle scattering problem in infinite
volume can be written in terms of spherical Bessel functions
$j_l(pr)$. For $pr\ll 1$, the Bessel functions behave as
$j_l(pr)\sim(pr)^l$. In a cubic box the lowest-possible momentum is
$p\sim 1/L$.  As the excited states are completely determined by the
nodal surface, which in our case is described by its radius
$r_\text{node}$, naive dimensional analysis suggests that we can
identify $r=r_\text{node}$. Hence, we expect that higher-$l$
contributions are suppressed by $(r_\text{node}/L)^l$ when comparing
to the leading contribution. Indeed, the coefficients $c_4$ and $c_6$,
shown in Fig.~\ref{Fig_Yldecomp}, are suppressed according to a power
law compared with the leading spherical contribution with $l=0$.

Although this is no longer true for $l>6$, the argument can still
serve as an estimate of the systematic uncertainty introduced through
the assumption of a spherical nodal surface in the AFDMC simulations
of the first excited state. A perturbative expansion of the energy in
terms of different $l$ contributions to the wave function,
\begin{equation}
E=\bra{\psi_{l=0}}H\ket{\psi_{l=0}}+c_4^2 \bra{\psi_{l=4}}H\ket{\psi_{l=4}} + \ldots \,,
\end{equation}
implies a correction proportional to $(c_4)^2$ when assuming that
$\bra{\psi_{l}}H\ket{\psi_{l}}$ is of the same order for all
$l$. Taking into account that the suppression seems to decrease with
higher $l$ we estimate $(c_4)^2$ conservatively as being of the order
of 1\% even though $c_4 \approx 0.045$ (see
Fig.~\ref{Fig_Yldecomp}). Therefore, an additional systematic error of
1\% is added to the statistical uncertainties from the QMC simulations
in Figs.~\ref{Fig_TP_EX} and~\ref{Fig_chiralLO_R1.0_EX}.

Furthermore, it is clear that the nodal surfaces shown in
Fig.~\ref{Fig_node} are less spherical for box sizes where the surface
is not closed.  This statement is supported by Fig.~\ref{Fig_Yldecomp}
where the coefficients for $L=30$~fm are larger than the other
contributions. This matches the deviations of AFDMC results with the spherical node from the L\"uscher predictions in Figs.~\ref{Fig_TP_EX}
and~\ref{Fig_chiralLO_R1.0_EX}, which are largest for $L=5,40,50$~fm.

Our diagonalization study suggests that a large improvement in our AFDMC results can be obtained by incorporating the first nonspherical contribution into the nodal surface
\begin{align}
	r_\text{node}(\theta,\phi)=c_{0} Y^c_{0}(\theta,\phi)+c_{4} Y^c_{4}(\theta,\phi) \,.
\end{align}
As discussed in Sec.~\ref{Sec_QMC} separate QMC runs were preformed on the two sides of the nodal surface in order to find an optimal set of parameters $c_0$ and $c_4$. Details on how the nonspherical nodal surface was incorporated into the Jastrow wave function can be found in Appendix~\ref{app:nodalsurf}.

As can be seen in Fig.~\ref{Fig_TP_EX}, the improved
nodal surface yields AFDMC results much closer to the L\"uscher
prediction.
However, one can see still some disagreement between the three methods
employed (diagonalization, AFDMC, and L\"uscher) at the smallest box
size considered, $L=5$~fm.
The results coming from the exact diagonalization and the AFDMC results
should agree well, as they do for larger box sizes.
That they do not suggests that our improved nodal surface is likely
missing higher-order $Y_l^c$ contributions and the associated uncertainties might be underestimated.
Since the L\"uscher results are based on the effective range expansion,
while the diagonalization uses the full potential,
deviations at large energies (corresponding to
small box sizes) are expected as soon as the effective range expansion is no longer accurate enough to
describe the phase shift.

\subsection{Chiral EFT interactions}
\label{Sec:chiralInt}

In this section, we present results for the different local chiral EFT
potentials from Ref.~\cite{Geze14long}.  To avoid large statistical
uncertainties, QMC simulations require interactions where all momentum
dependencies up to quadratic terms can be separated.  This requirement
is met by local potentials~\cite{Carl15RMP}.  However, chiral EFT
interactions are usually formulated in momentum space and are
typically nonlocal.  Local chiral $NN$ potentials have been developed
recently up to next-to-next-to-leading order (N$^2$LO) in the chiral
power counting and applied in calculations of neutron matter, light
nuclei, and neutron-alpha scattering using continuum QMC
methods~\cite{Geze13QMCchi,Geze14long,Lynn14QMCln,Tews20163N,Lynn20163N}.

Table~\ref{tab:gfmcafdmc} compares results for the leading chiral potential for both GFMC and AFDMC methods. As discussed before, uncertainty estimates are very similar and we limit our plots to results from the AFDMC method.

The range of the chiral potentials exceeds the range of the contact potential in Eq.~\eqref{Eq_testpot}. 
A check for convergence shows that, for box sizes up to L = 20 fm, inclusion
of copies of the original box up to the second-to-nearest is
required to reach truncation uncertainties comparable to the statistical errors, while beyond L = 20 fm at most the nearest copies need to be included.

\begin{figure}[t]
\includegraphics[width=0.48\textwidth,clip=]{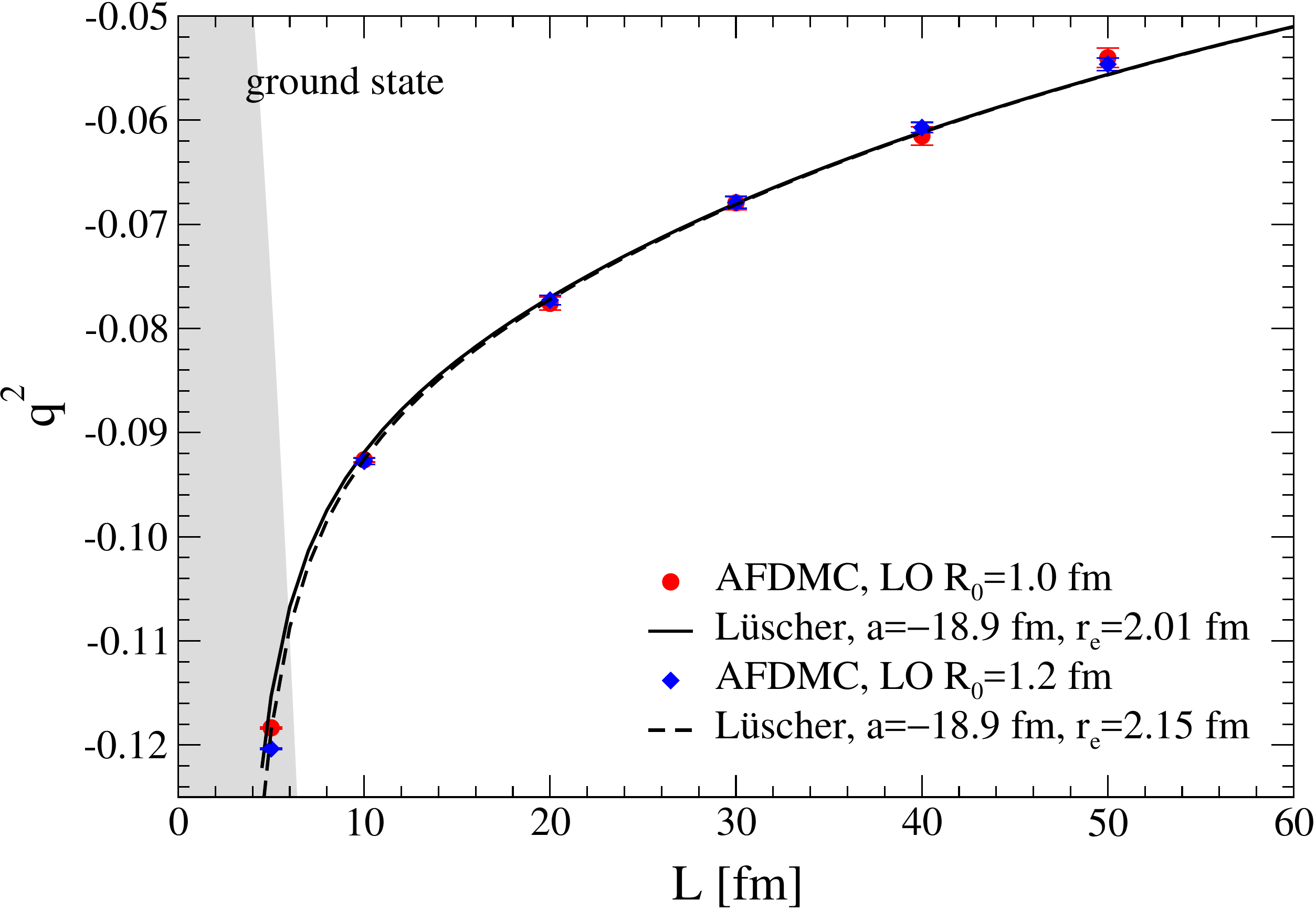}
\caption{(Color online) AFDMC results for the energy of two neutrons 
in the ground state in finite volume with the LO chiral EFT interaction
compared with the L\"uscher formula for different box sizes~$L$.  The
cutoffs $R_0=1.0$~fm (red circles and solid line) and $R_0=1.2$~fm (blue
diamonds and dashed line) are used.  The energies are given in terms of
the dimensionless quantity $q^2=E ML^2/(4\pi^2)$.  The region where
$|p|> m_\pi/2$ is indicated by the gray band; see
Sec.~\ref{Sec_Luescher}.}
\label{Fig_chiralLO}
\end{figure}

\begin{figure}[t]
\includegraphics[width=0.48\textwidth,clip=]{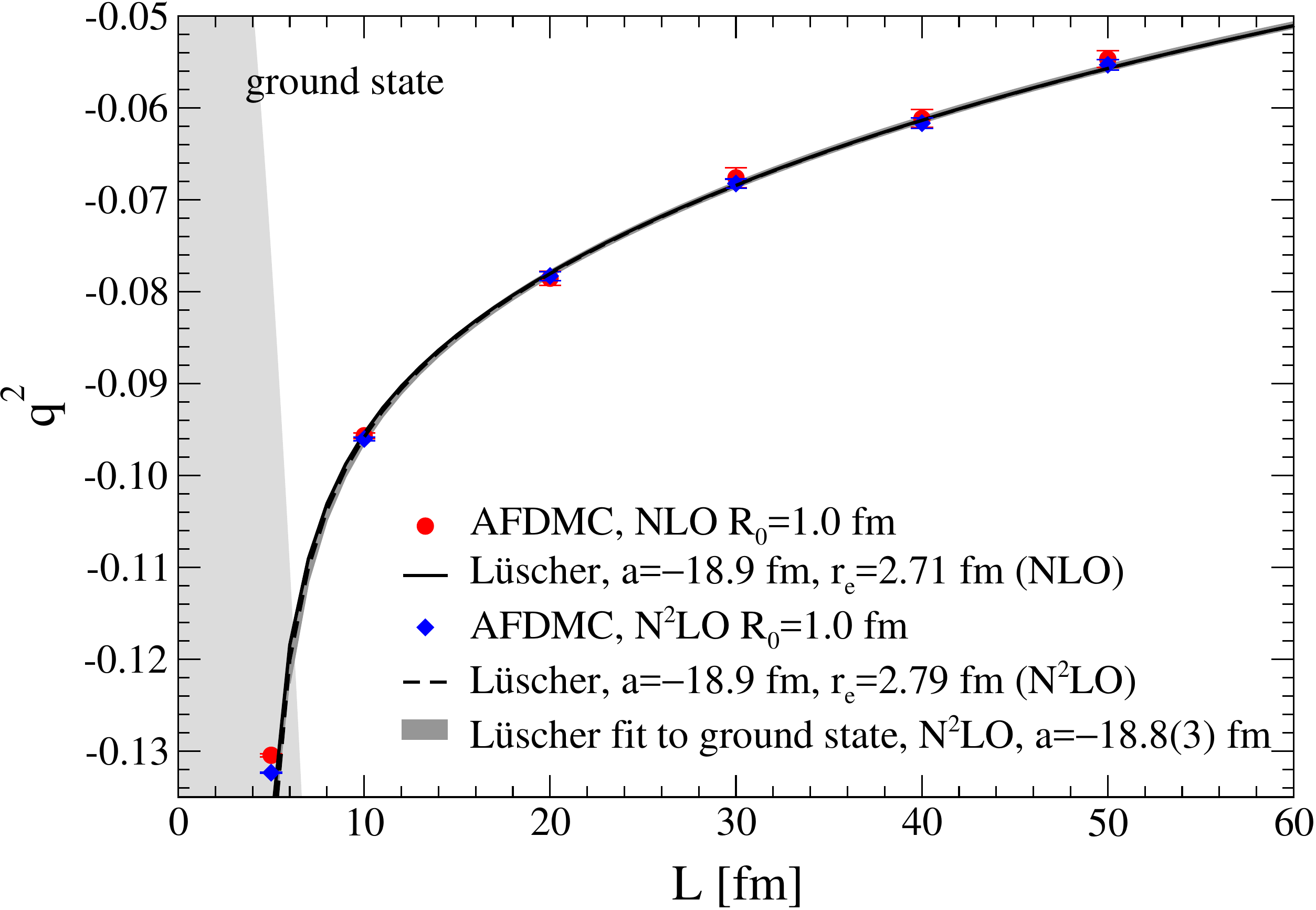}
\caption{(Color online) AFDMC results for the energy of two neutrons 
in the ground state in finite volume with the NLO and N$^2$LO chiral
EFT interactions with cutoff $R_0=1.0$~fm compared with the L\"uscher
formula for different box sizes $L$.  The results at NLO (N$^2$LO) are
given as the red circles and solid line (blue diamonds and dashed line).  The
dark gray band shows a fit (as described in the text) to the AFDMC
results for the N$^2$LO chiral potential.  The energies are given in
terms of the dimensionless quantity $q^2=E ML^2/(4\pi^2)$.  Points in
the region $|p|> m_\pi/2$ indicated by the gray band are not included in
the fit; see Sec.~\ref{Sec_Luescher}.}
\label{Fig_chiralNLO_R1.0}
\end{figure}

\begin{figure}[t]
\includegraphics[width=0.48\textwidth,clip=]{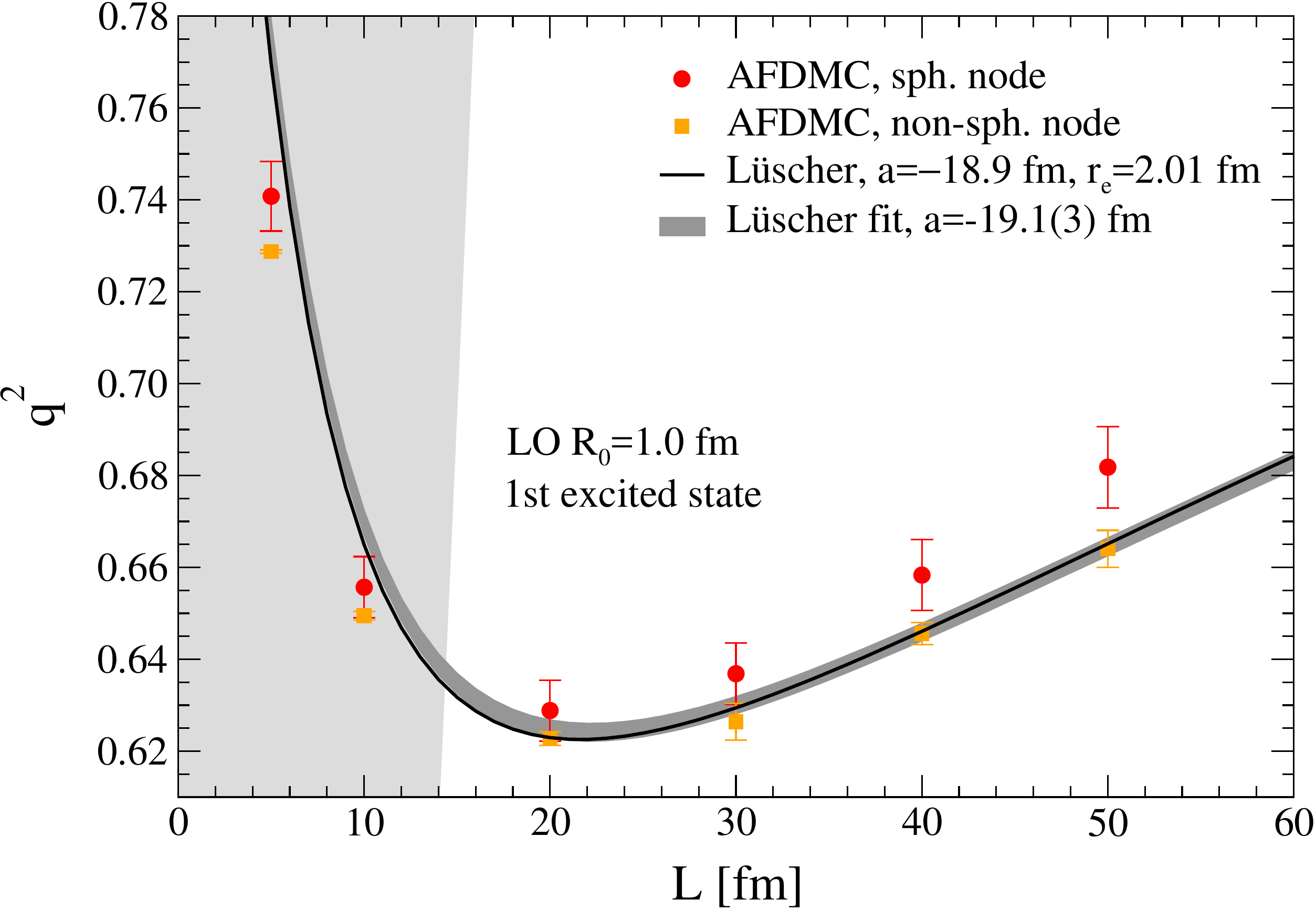}
\caption{(Color online) AFDMC results for the energy of two neutrons 
in the first excited state in finite volume with the LO chiral EFT
interaction with cutoff $R_0=1.0$~fm (red circles) compared with the
L\"uscher formula (solid line) for different box sizes $L$.  The error
bars on the AFDMC results with a spherical nodal surface include both statistical uncertainties and a
systematic uncertainty of 1~\% discussed in the text in
Sec.~\ref{Sec:Diag}.  The dark gray band shows a combined fit (as
described in the text) to the ground- and first-excited-state AFDMC
results for the LO chiral potential.  The energies are given in terms
of the dimensionless quantity $q^2=E ML^2/(4\pi^2)$.  Points in the
region $|p|> m_\pi/2$ indicated by the gray band are not included in the
fit; see Sec.~\ref{Sec_Luescher}.}
\label{Fig_chiralLO_R1.0_EX}
\end{figure}

In Fig.~\ref{Fig_chiralLO} we show results of AFDMC simulations which
were performed using the chiral leading-order (LO) potential for
$R_0=1.0$~fm and $R_0=1.2$~fm, corresponding to cutoffs of 500 MeV and
400 MeV in momentum space, respectively.  The corresponding scattering
lengths and effective ranges were obtained by calculating phase shifts
in the infinite volume.  Similar to the previous cases we compare
L\"uscher results by using the scattering parameters to AFDMC results for
different box sizes.  Figure~\ref{Fig_chiralNLO_R1.0} shows results
for the chiral NLO and N$^2$LO potentials for $R_0=1.0$~fm.  In all
cases the overall agreement for box sizes $L\geqslant 10$~fm is
excellent, while the AFDMC results for ground-state energies at
$L=5$~fm start to deviate from the L\"uscher prediction. 
In Fig.~\ref{Fig_chiralLO_R1.0_EX} results for the excited states of the chiral LO potential for both a spherical nodal surface as described in Sec.~\ref{Sec_QMC} and a nonspherical nodal surface as introduced in Sec.~\ref{Sec:Diag} are shown. While the spherical node over-predicts at large $L$, the improved nodal surface yields results consistent with the L\"uscher prediction in this region.
However, in this context it is worth pointing out that for the chiral potentials the nodal surfaces tend to be more deformed than for the contact potential, which implies that the uncertainties for the small boxes are likely underestimated.

As discussed in Sec.~\ref{Sec_Luescher} the
analytic continuation of the L\"uscher formula is limited to the
threshold of pionless EFT $|p|<m_\pi/2$.  Figures~\ref{Fig_chiralLO} -- %, Fig_chiralNLO_R1.0,
\ref{Fig_chiralLO_R1.0_EX} show the
corresponding maximal value for $q^2$.  The absolute values of the
AFDMC energies for $L=5$~fm (ground states) and $L=5,10$~fm (excited
state) for the different chiral potentials exceed the threshold of
pionless EFT, and hence the (exponentially suppressed) disagreement is to be expected.  
In the end, this effect reflects the necessity of including pions in the effective theory for
the correct description of processes where momenta are of the order of
the pion mass. However, we find that the size of the corrections is smaller than naively expected:
for the smallest box size with $m_\pi L=3.5$ the leading effect should scale as
$c_1e^{-m_\pi L}=18\%$~\cite{Sato:2007ms}, where $c_1=6$ denotes the multiplicity of nearest
neighbors, but the actually observed deviation merely amounts to about $3\%$.  
This finding could be related to the observation in~\cite{Sato:2007ms} that, for 
a realistic $NN$ potential, the effective scale in the exponent can exceed the pion mass,
leading to a stronger suppression than expected from the one-pion exchange alone.

\begin{figure}[t]
\includegraphics[width=0.48\textwidth,clip=]{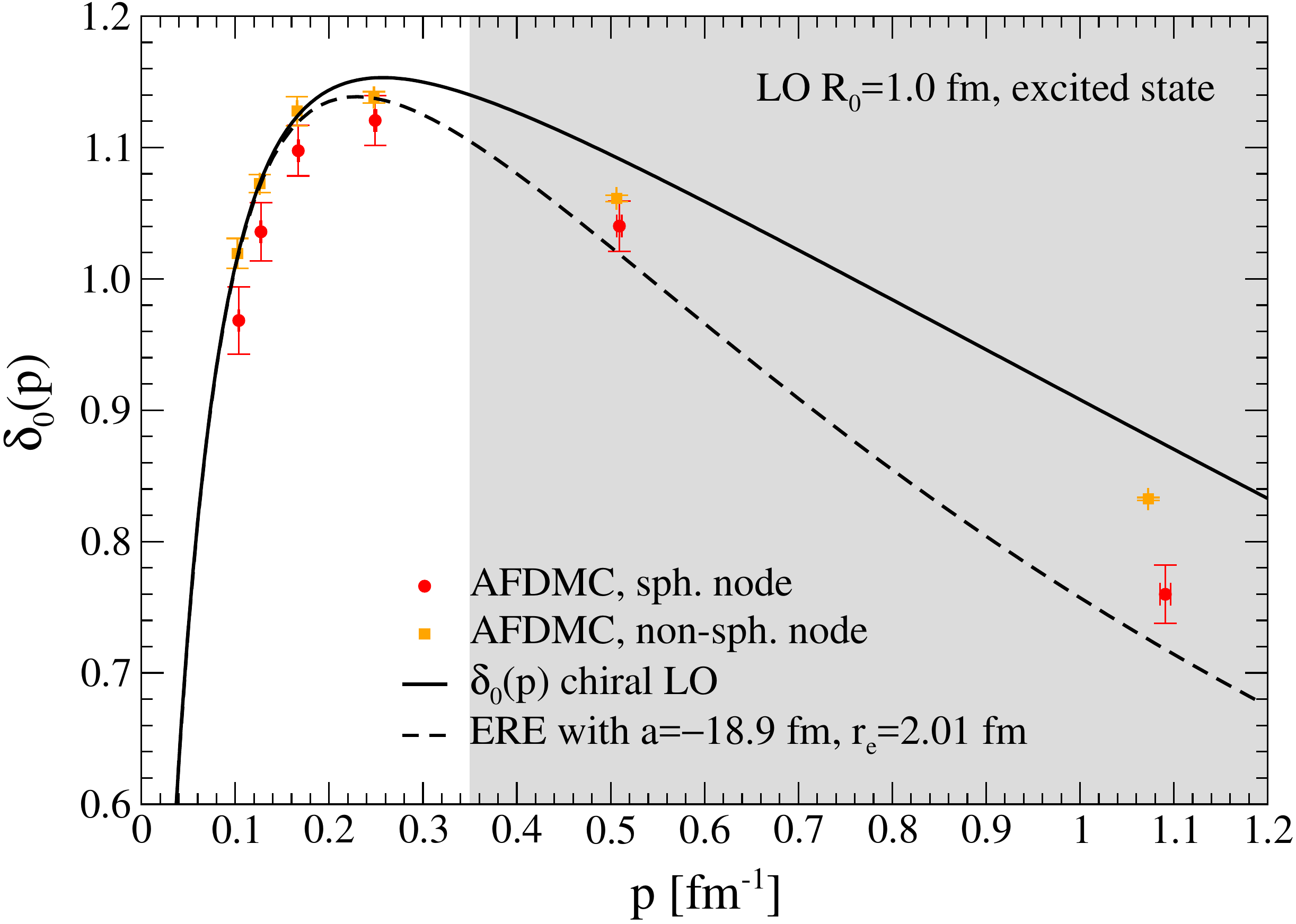}
\caption{(Color online) $^1S_0$ $nn$ phase shift $\delta_0(p)$ from 
AFDMC results for the first excited state in finite volume with the LO
chiral EFT interaction with cutoff $R_0=1.0$~fm with spherical nodal surface (red circles) and nonspherical nodal surface (orange squares) compared
with the phase shift obtained from $nn$ scattering in infinite volume
(solid line).  The error bars on the AFDMC results include both
statistical uncertainties and a systematic uncertainty of 1~\%
discussed in the text in Sec.~\ref{Sec:Diag}. The dashed line shows the phase shift obtained from the effective range expansion (ERE) with $a=-18.9$~fm and $\re=2.01$~fm. The region where $|p|>
m_\pi/2$ is indicated by the gray band; see Sec.~\ref{Sec_Luescher}. }
\label{Fig_chiralLO_phaseshift}
\end{figure}

Figure~\ref{Fig_chiralLO_phaseshift} shows phase shifts obtained by
solving Eq.~\eqref{Eq_Luescher} for the AFDMC results for the excited
state with the LO chiral potential shown in
Fig.~\ref{Fig_chiralLO_R1.0_EX} (red circles and orange squares).  A direct extraction of phase shifts
from finite-volume energies is only possible for states with $E>0$ as
only the effective-range expansion provides an analytic continuation
to imaginary momenta corresponding to bound states.  We compare the AFDMC results to the
phase shifts obtained by solving the $nn$ scattering problem for the same
chiral LO potential in infinite volume (black line). As in
Fig.~\ref{Fig_chiralLO_R1.0_EX}, the overall trend is correctly
reproduced by both spherical and nonspherical node results. We show again in gray the region for which momenta exceed
the regime of pionless EFT, $|p|>m_\pi/2$. The AFDMC data with a spherical node underestimate
the phase shift over the whole region. The improved nodal surface yields phase shifts in very good agreement with the infinite-volume phase shift at small momenta. Beyond the regime of pionless EFT the results are still too low but are significantly closer to the phase shift than the spherical node results. 
We also show the phase shift obtained from the effective range expansion with the first two parameters $a=-18.9$~fm and $\re=2.01$~fm (dashed line), as used for the L\"uscher result in Fig.~\ref{Fig_chiralLO_R1.0_EX}. 
For momenta above the
strict range of validity, the AFDMC results are larger than the phase shift from the truncated effective range expansion. This corresponds to Fig.~\ref{Fig_chiralLO_R1.0_EX} where these points lie below the L\"uscher result. 

\section{Conclusions}
\label{Sec_conclusions}

In this work, we present first results for the two-neutron
finite-volume ground and first-excited states using AFDMC, both for a
contact potential and chiral EFT interactions.  To extract the
excited-state energy we used an approximate method based on an
iterative determination of the nodal surface of the excited-state wave
function, with systematic uncertainties estimated by direct
diagonalization of the contact-potential Hamiltonian. Having obtained the exact nodal surfaces through the diagonalization, we incorporated nonspherical nodal surfaces in the AFDMC method, which significantly improves results for the excited states. Using
L\"uscher's method to extract the scattering length and effective
range from fits to the finite-volume energy levels of the ground and
excited state, we found good agreement with the scattering parameters
determined directly in infinite volume, which demonstrates the
viability of the AFDMC approach for the calculation of the
finite-volume two-particle spectrum.

In this way, our work establishes QMC techniques as powerful methods
for the matching of chiral EFT and lattice QCD results. By equating
the finite-volume energy levels one can directly determine the LECs in
the chiral potential without the necessity of first extracting the
infinite-volume phase shift.  Advantages of this procedure concern the
fact that it evades limitations of the L\"uscher formula for small
volumes and should generalize straightforwardly to the multi-body
system, to avoid the complexity typically inherent in the extension of
L\"uscher's approach beyond the two-body sector.  For the extraction of resonance properties along these lines, we anticipate control over excited states to be essential. In this regard, the recovery of the virtual state in the two-neutron system as reflected by the large scattering length, both in the ground and first excited state, can be considered a successful proof of principle.

\begin{acknowledgments}

We thank J.\ A.\ Carlson, Z.\ Davoudi, A.\ Rusetsky, M.\ J.\ Savage,
K.\ E.\ Schmidt, S.\ R.\ Sharpe, and K.\ A.\ Wendt for helpful discussions. This work was supported in
part by the ERC Grant No.~307986 STRONGINT, the Deutsche
Forschungsgemeinschaft through Grant SFB 1245, the US Department of
Energy Grant Nos.~DE-AC52-06NA25396 and DE-FG02-00ER41132, the
National Science Foundation Grant No.~PHY-1430152 (JINA-CEE), the
NUCLEI SciDAC program, the LANL LDRD program, and the Natural Sciences
and Engineering Research Council of Canada. The computations were
performed at the J\"ulich Supercomputing Center as well as at NERSC,
which is supported by the US Department of Energy Contract
No.~DE-AC02-05CH11231.
We thank the Institute for Nuclear Theory at the University of Washington for
its hospitality and the Department of Energy for partial support during the
completion of this work.

\end{acknowledgments}

\appendix

\section{Evaluation of $\boldsymbol{S(\eta)}$}
\label{app:Seta}

The definition of $S(\eta)$ that appears in Eq.~\eqref{S_eta_loop} can
be evaluated numerically; however, in practice it converges relatively
slowly.  A more efficient approach used in Ref.~\cite{Lues86long}
relies on Poisson's summation formula as well as the fact that
$S(\eta)$ equals the analytic continuation of
\begin{equation}
Z_{00}(s,\eta)=\sum_{{\bf j}}\frac{1}{({\bf j}^2-\eta)^s} \,, \qquad 
\Re s>\frac{3}{2}\,,
\end{equation}
for $s\to 1$, which leads to
\begin{align}
S(\eta)=&\sum_{{\bf j}^2<\eta}\frac{1}{{\bf j}^2-\eta}+\int_0^1 dt\, F^1_{00}(t,\eta)\nonumber\\
&+\int_1^\infty dt\, F_{00}(t,\eta)+\sum_{i=0}^1\biggl(\frac{A_i}{i+1}+\frac{B_i}{i-\frac{1}{2}}\biggr)\,,
\label{Eq:S_numerical}
\end{align}
where
\begin{align}
F^1_{00}(t,\eta)=&-\sum_{{\bf j}^2\leqslant\eta}e^{t(\eta-{\bf j}^2)}+\biggl(\frac{\pi}{t}\biggr)^{3/2}e^{t\eta}\sum_{{\bf j}} e^{-\frac{\pi^2}{t}{\bf j}^2}\notag\\
&-\sum_{i=0}^1(A_it^i+B_it^{i-3/2})\,,\notag\\
F_{00}(t,\eta)=&\sum_{{\bf j}^2>\eta}e^{-t({\bf j}^2-\eta)}\,,\notag\\
A_i=&-\frac{1}{i!}\sum_{{\bf j}^2\leqslant\eta}(\eta-{\bf j}^2)^i\,,\notag\\
B_i=&\pi^{3/2}\frac{\eta^i}{i!}\,.
\end{align}
This representation accelerates convergence exponentially and can be
easily implemented by using standard integration routines.

\section{Jastrow wave function for nonspherical nodal surfaces}
\label{app:nodalsurf}
The Jastrow wave function $\psi_J(r)$ commonly used in QMC simulations is given by the solutions of the radial Schr\"odinger equation with the central part of the potential. The solutions are required to meet the following boundary conditions:
\begin{align}
	\psi_J(0)&=u_0\, ,\notag\\
	\psi_J(L/2)&=1\, ,\notag\\
	\psi_J'(0)&=0\, ,\notag\\
	\psi_J'(L/2)&=0\, ,
	\label{Eq:nonsphcond}
\end{align}
where $u_0$ is a constant.
Furthermore, for the ground-state trial wave function it is required that there be no nodes in $\psi_J(r)$.

The spherical nodal surface was implemented by constructing a Jastrow function with a single node. This was achieved by writing the Jastrow in terms of a sum of different solutions of the radial Schr\"odinger equation
\begin{align}
	\psi_J^\text{sph}(r)=N(\psi_J^1(r)-c \;\psi_J^0(r))\,,
\end{align}
where $N$ denotes a normalization constant and the superscript in $\psi_J^i(r)$ denotes the number of nodes. By changing the parameter $c$ it is possible to adjust the position $r_\text{node}$ of the node such that $\psi_J^\text{sph}(r_\text{node})=0$.

To improve the nodal surface in the QMC method we take advantage of the analysis of the nodal surface obtained from the diagonalization in Sec.~\ref{Sec:Diag}. Usually the Jastrow function is a radial function only allowing for spherical nodal surfaces. If the nonspherical nodal surface is to be reproduced by the Jastrow function, angular dependencies have to be introduced.

Including the first nonspherical contribution in the nodal surface corresponds to adding the cubic harmonic with $l=4$ to the spherical term:
\begin{align}
	r_\text{node}(\hat{\bf r})=c_0 Y^c_0(\hat{\bf r})+c_4 Y^c_4(\hat{\bf r})\,,
\end{align}
where $Y^c_l$ denote cubical harmonics and $\hat{\bf r}={\bf r}/r$ is the unit vector pointing in the direction of ${\bf r}$. $c_0$ and $c_4$ are coefficients defining the nodal surface.

The function defined by
\begin{align}
	f_\text{non-sph}({\bf r})=\psi_J^1(r)-\frac{\psi_J^1(r_\text{node}(\hat{\bf r}))}{\psi_J^0(r_\text{node}(\hat{\bf r}))}\psi_J^0(r)
	\label{Eq:jasnonsph}
\end{align}
vanishes when $r=r_\text{node}$ for a given direction $\hat{\bf r}$. However, this function does not meet the boundary conditions in Eq.~\eqref{Eq:nonsphcond}. Furthermore it is not continuous at $r\rightarrow0$ since for vectors ${\bf r}_1$ and ${\bf r}_2$ pointing in different directions
\begin{align}
	\lim_{r_1\rightarrow 0} f_\text{non-sph}({\bf r}_1)\neq \lim_{r_2\rightarrow 0}f_\text{non-sph}({\bf r}_2)\,.
\end{align}
Therefore, the function defined in Eq.~\eqref{Eq:jasnonsph} is multiplied by a normalizing function
\begin{align}
	n({\bf r})&=n_3(\hat{\bf r}) r^3 +n_2(\hat{\bf r}) r^2 + n_1(\hat{\bf r}) r + n_0(\hat{\bf r})\,,\notag\\
	n_3(\hat{\bf r})&=\frac{16}{L^3}\Bigl(\frac{u_0}{a(\hat{\bf r})}-\frac{1}{b(\hat{\bf r})}\Bigr)\,,\notag\\
	n_2(\hat{\bf r})&=\frac{12}{L^2}\Bigl(\frac{1}{b(\hat{\bf r})}-\frac{u_0}{a(\hat{\bf r})}\Bigr)\,,\notag\\
	n_1(\hat{\bf r})&=0\,,\notag\\
	n_0(\hat{\bf r})&=\frac{u_0}{a(\hat{\bf r})}\,,
\end{align}
where
\begin{align}
	a(\hat{\bf r})&=f_\text{non-sph}({\bf r})|_{r=0}\,,\notag\\
	b(\hat{\bf r})&=f_\text{non-sph}({\bf r})|_{r=L/2}\,,
	\end{align}
and $u_0<0$ defines the value of the Jastrow at $r=0$. The nonspherical Jastrow is then given by
\begin{align}
	\psi_J^\text{non-sph}({\bf r})=f_\text{non-sph}({\bf r})n(\hat{\bf r})\,,
\end{align}
which obeys the required conditions in Eq.~\eqref{Eq:nonsphcond}. Now, the excited-state energies can be found as discussed in Sec.~\ref{Sec_QMC} by adjusting the parameters $c_0$ and $c_4$.

\bibliography{localbib}{}
\bibliographystyle{apsrev}

\end{document}